# Quantum Circuit Parameters Learning with Gradient Descent Using Backpropagation

Masaya Watabe[1], Kodai Shiba[1,3,] Masaru Sogabe[3], Katsuyoshi Sakamoto[1,2,] Tomah Sogabe[1,2,3*]

[1] Engineering department, The University of Electro-Communications, Tokyo, Japan
[2] i-PERC, The University of Electro-Communications, Tokyo, Japan
[3] Grid, Inc. Tokyo, Japan

E-mail: sogabe@uec.ac.jp

**Abstract**

Quantum computing has the potential to outperform classical computers and is expected to play an active role in various fields. In quantum machine learning, a quantum computer has been found useful for enhanced feature representation and high dimensional state or function approximation. Quantum-Classical hybrid algorithms are proposed in recent years for this purpose under the Noisy-Intermediate Scale Quantum computer (NISQ) environment. Under this scheme, the role played by classical computer is the parameter tuning, parameter optimization, and parameter update for the quantum circuit. In this paper, we propose a gradient descent based backpropagation algorithm that can efficiently calculate the gradient in parameter optimization and update the parameter for quantum circuit learning, which outperforms the current parameter search algorithms in terms of computing speed while presents the same or even higher test accuracy.

## 1. Introduction

. There are the famous Di Vincenzo criteria for the question "What are the standards for quantum computers" or "What are the elements necessary for a" real "quantum computer?" [1]. Di Vincenzo criteria contain seven criteria, and each one is steadily cleared by the development of technology in recent years, for example, qubits and quantum gates have increased from tens to thousands with the development of superconducting technology, spin control technology, and microwave resonance technology. The Di Vincenzo criteria are essential for all items; still, the most difficult thing from the results of recent research and development is the third criterion, "Coherence time continues until the quantum computation is completed". This condition has a deep meaning, and it can be said that it is a life-and-death problem of a quantum computer that is questioned by the completeness of physical conditions under which quantum superposition and quantum entanglement are the critical elements of a quantum computer.

Quantum coherence refers to a coordinated and precise movement of qubits. However, in reality, qubits are very fragile and are subject to errors due to the phenomenon of decoherence. Quantum coherence error is the biggest reason why it is challenging to realize quantum computers. It is difficult to precisely control the quantum state such as spin as expected, and bit flip and phase inversion occur due to the fluctuation in the surrounding environment and noise. Such an error is called a quantum error. The amount of quantum errors that potentially exist or the number of quantum errors that occur in a normal stochastic qubit is an important parameter that affects quantum computers, and it has been actively studied in recent years. Quantum computer, which possesses considerable quantum errors, is called the Noisy-Intermediate Scale Quantum computer (NISQ) [2]. Under the NISQ circumstance, it is necessary to develop fault-tolerant quantum computation methods that provide error resilience. There are two solutions to this problem. One is to perform quantum computing while correcting quantum errors in the presence of errors. Another approach is to develop a hybrid quantum-classical algorithm which completes the quantum computing before the quantum error becomes fatal and shifts the rest of the task to the classical computer when a severe quantum error occurs. The latter approach has triggered many algorithms, such as quantum approximation optimization algorithm (QAOA) [3] and variation quantum eigensolver (VQE) [4] and many others[5–7]. The quantum-classical algorithm aims to seek the 'quantum advantage' rather than 'quantum supremacy' [8]. Quantum supremacy states that quantum computer must prove that it can achieve a level either in terms of speed or solution finding, which can never be completed by the classical computer. It has been considered that the quantum supremacy may appear in several decades later and 'quantum supremacy' reported so far are either overstating or lack of fair comparison [9,10]. From this point of view, the quantum advantage is a more realistic goal, and it aims to find the concrete and beneficial applications of the NISQ quantum computer in the near future. Within the scope of quantum advantage, application of quantum computer can be expanded far beyond computing speed racing to be used in various fields such as representing wavefunction in quantum chemistry field [11–15], or working as a quantum kernel to represent enhanced high dimensional feature in the field of machine learning [16–19].

A quantum-classical hybrid algorithm needs to build an efficient simulation channel to connect 'the quantum and the classic' organically. In QAOA, VQE, or other hybrid NISQ algorithms, there exists a challenging task to optimize the model parameter. In all these algorithms, the parameter search and updating are performed in the classical computer. In a complete classical approach, the optimal parameter search is usually categorized as a mathematical optimization problem, where various methods both gradient-based and non-gradient based have been widely utilized. For the quantum circuit learning, so far, most of parameter searching algorithm is based on non-gradient ones such as Nelder–Mead method [20]. However, recently, gradient-based ones such as SPSA [21] and a finite difference method have been reported [22].

In this article, we propose an error backpropagation algorithm on quantum circuit learning to calculate the gradient required in parameter optimization efficiently. The purpose of this work is to develop a gradient-based circuit learning algorithm with superior learning speed to the ones reported so far. The error backpropagation method is known as an efficient method for calculating gradients in the field of deep neural network machine learning when updating parameters using the gradient descent method [23]. Further speed improvement can be easily realized through using the GPGPU technique, which is again well established and under significant growth in the field of deep learning[24].

The idea behind our proposal is described as follows. By carefully examining the simulation process of a quantum circuit, if the input quantum state is $|\psi_{in}\rangle$ and a specific quantum gate $U(\theta)$ is applied as shown in Fig.1, the output state $|\psi_{out}\rangle$ can be expressed by the dot product of the input state and quantum gate.

$$|\psi_{out}\rangle = U(\theta)|\psi_{in}\rangle \quad (1)$$

On the other hand, the calculation process of a fully connected neural network without activation function can be written as $\mathbf{Y} = \mathbf{W} \cdot \mathbf{X}$, where $\mathbf{X}$ is the input vector, $\mathbf{W}$ is the weight matrix of the network, and $\mathbf{Y}$ is the output. It can be seen that the quantum gate $U(\theta)$ is very similar to the network weight matrix $\mathbf{W}$. This shows that backpropagation algorithms that are used for deep neural networks can be modified to some extent to be applied to the simulation process of quantum circuit learning.

The method we proposed makes it possible to reduce the time significantly for gradient calculation when the number of qubits is increased, or the depth of the circuit (the number of gates) is increased. Meanwhile, by taking advantage of GPGPU, which is developed upon backpropagation, it is expected that using gradient-based backpropagation in the NISQ hybrid algorithm will further facilitate parameter search when many qubits and deeper circuits are deployed.

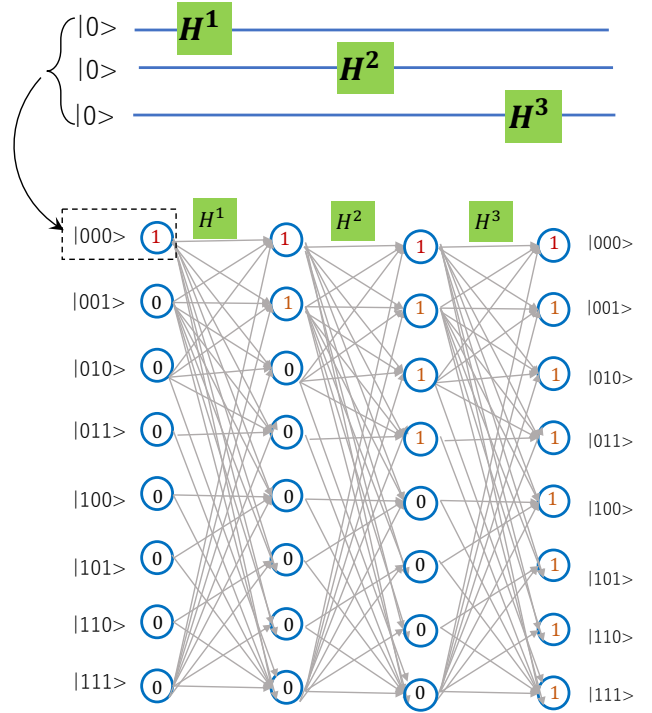

Fig. 1 Example of three gates quantum circuit and its corresponding fully connected quantum network, showing similarity to a four-layer-neural network with a equal number of nodes in the input layer, middle layer and output layer. Note that the amplitude value is not normalized for better eye-guiding illustration.

## 1.1 Quantum Backpropagation Algorithm

As shown in Fig.1, a quantum circuit can be effectively represented by a fully connected quantum network with significant similarity to the conventional neural network except for the two facts: 1) there is no activation function applied upon each node, so the node is not considered as a neuron (or assuming an identical activation function); 2) the number of nodes are equal among all the input layer, middle layer as well as output layer, so there is no notion of 'hidden' layer since the dimensionality of each layer is the same. Noticed that the states shown as input in the quantum circuit is only one of the $2^n$ ($n$ is the number of qubits) with the amplitude of '1' (not normalized), see Fig.1 for details. The network similarity implies that the learning algorithm, such as the backpropagation heavily used in the field of deep machine learning, can be shared by the quantum circuit as well.

In general, the backpropagation method uses the chain rule of a partial differential to propagate the gradient back from the network output and calculate the gradient of the weights. Owing to the chain rule, the backpropagation can be done only at the input/output relationship at the computation cost of a node[23]. In the simulation of quantum computing by error backpropagation, the quantum state $|\psi\rangle$ and the quantum gates are represented by complex value. Here we will show the derivation details regarding the quantum backpropagation in complex-valued vector space. When the input of n qubits is $|\psi_{in}\rangle$ and the quantum circuit parameter network $W(\theta)$ is applied, the output $|\psi_{out}\rangle$ can be expressed as:

$$W(\theta)|\psi_{in}\rangle = \sum_{j=0}^{2^n-1} c_\theta^j |j\rangle = |\psi_{out}\rangle \quad (2)$$

where $c_\theta^j$ is the probability amplitude of state $|j\rangle$ and $|c_\theta^j|^2 = p_\theta^j$ is the observation probability of state $|j\rangle$. If loss function $L$ can be expressed by using observation probability determined by quantum measurement, the gradient of the learning parameter can be described as:

$$\frac{\partial L}{\partial \theta} = \frac{\partial L}{\partial p_\theta^j} \cdot \frac{\partial p_\theta^j}{\partial \theta} \quad (3)$$

Since

$$p_\theta^j = |c_\theta^j|^2 = c_\theta^j \overline{c_\theta^j} \quad (4)$$

where $\overline{c_\theta^j}$ is the conjugate of $c_\theta^j$, Therefore, the gradient of observation probability can be further expanded as:

$$\frac{\partial p_\theta^j}{\partial \theta} = \frac{\partial c_\theta^j \overline{c_\theta^j}}{\partial \theta} = \overline{c_\theta^j} \frac{\partial c_\theta^j}{\partial \theta} + c_\theta^j \frac{\partial \overline{c_\theta^j}}{\partial \theta} \quad (5)$$

Formula (5) can be further expanded as:

$$\overline{c_\theta^j} \frac{\partial c_\theta^j}{\partial \theta} + c_\theta^j \frac{\partial \overline{c_\theta^j}}{\partial \theta} = \overline{c_\theta^j} \frac{\partial c_\theta^j}{\partial \theta} + \overline{\overline{c_\theta^j} \frac{\partial c_\theta^j}{\partial \theta}} \quad (6)$$

Formula (6) contains complex value but can be nicely summed out as real value shown below:

$$\overline{c_\theta^j} \frac{\partial c_\theta^j}{\partial \theta} + \overline{\overline{c_\theta^j} \frac{\partial c_\theta^j}{\partial \theta}} = 2\mathrm{Re}\left[\overline{c_\theta^j} \frac{\partial c_\theta^j}{\partial \theta}\right] \quad (7)$$

Using the formula $\frac{\partial p_\theta^j}{\partial c_\theta^j} = \overline{c_\theta^j}$, the $\overline{c_\theta^j}$ can be replaced as follows:

$$\overline{c_\theta^j} \frac{\partial c_\theta^j}{\partial \theta} + \overline{\overline{c_\theta^j} \frac{\partial c_\theta^j}{\partial \theta}} = 2\mathrm{Re}\left[\frac{\partial p_\theta^j}{\partial c_\theta^j} \frac{\partial c_\theta^j}{\partial \theta}\right] \quad (8)$$

Therefore,

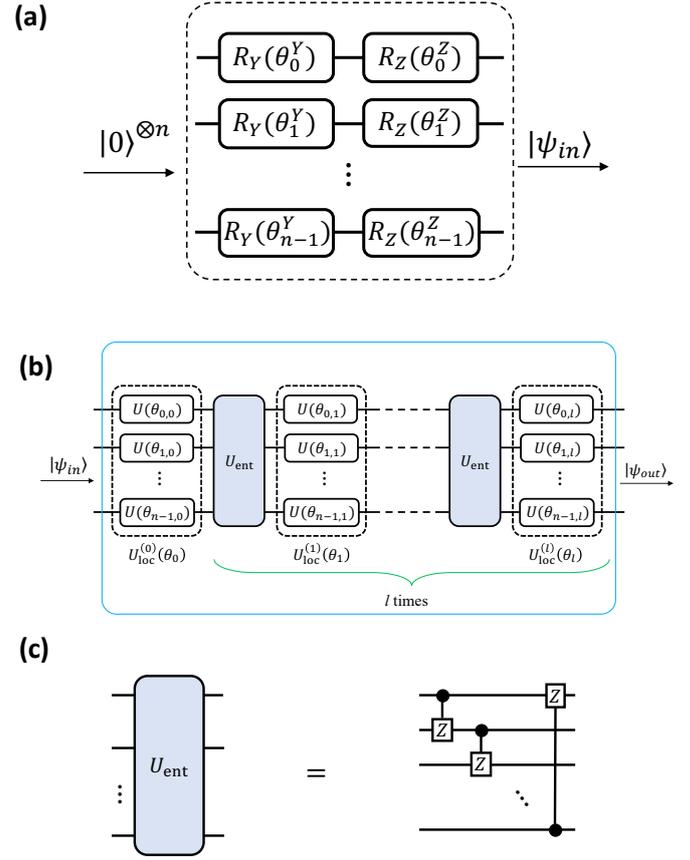

Fig. 2 (a) Preparation of input state by a unitary input gate $U_{in}(x)$ exemplified by a series of rotation gates. (b) Quantum circuit to present variational parameter state $W(\theta)$. $l$ denotes the depth of quantum circuit. (c) Quantum entanglement circuit where $U_{ent}$ gate is composed of CZ gates from qubit j to qubit (j+1) mod n, $j \in \{0, \ldots, n-1\}$.

$$\frac{\partial L}{\partial \theta} = 2\mathrm{Re}\left[\frac{\partial L}{\partial p_\theta^j} \frac{\partial p_\theta^j}{\partial c_\theta^j} \frac{\partial c_\theta^j}{\partial \theta}\right]. \quad (9)$$

$\frac{\partial L}{\partial p_\theta^j} \frac{\partial p_\theta^j}{\partial c_\theta^j} \frac{\partial c_\theta^j}{\partial \theta}$ can be obtained by error backpropagation in the same way as the conventional calculation used in deep neural network[25]. Meanwhile, one advantage of the proposed method is that the quantum gate matrix containing complex value is converted to real value. The gradient of the loss function with respect to $\theta$ can be obtained from the real part of the value of the complex vector space calculated by the conventional backpropagation. More detailed derivation regarding backpropagation at each node using a computation graph is given in the Appendix for reference.

## 2. Experiment and results

Next, we conducted the experiment with the supervised learning tasks, including both regression and classification problems, to verify the validity of the proposed quantum backpropagation algorithm.

The quantum circuit consists of a unitary input gate $U_{in}(x)$ that creates an input state from classical input data $x$ and a unitary gate $W(\theta)$ with parameters $\theta$. We use $U_{in}(x) = \otimes_{j=0}^{n-1} R_Z(\theta_j^Z) R_Y(\theta_j^Y)$ as proposed in reference [22] a unitary input gate, as shown in Fig. 2(a). We use $W(\theta) = U_{loc}^{(l)}(\theta_l) U_{ent} \cdots U_{loc}^{(1)}(\theta_1) U_{ent} U_{loc}^{(0)}(\theta_0)$ as proposed in reference [26], therefore $U_{loc}^{(k)}(\theta_k) = \otimes_{j=0}^{n-1} U(\theta_{j,k})$, shown in Fig.2(b). The layer $U_{loc}^{(k)}(\theta_k)$ is comprised of local single qubit rotations. We only use $Y$ and $Z$ rotations. So, $U(\theta_{j,k}) = R_Z(\theta_{j,k}^Z) R_Y(\theta_{j,k}^Y)$. Each $\theta$ is parameterized $\theta_k \in \mathbb{R}^{2n}, \theta_{j,k} \in \mathbb{R}^2$. $U_{ent}$ is entangling gates. We use controlled-$Z$ gates ($CZ$) as $U_{ent}$. The overall quantum circuit is shown in Fig.2(c):

## 2.1 Regression

In regression tasks, the circuit parameters were set to n = 3 and l = 3; that is to say, the number of qubits is 3 and the depth of the circuit is 4. The expected value of observable Pauli Z for the first qubit was obtained from the output state $|\psi_{out}\rangle$ of the circuit. One-dimensional data $x$ is input by setting circuit parameters as

$$\theta^Z = \cos^{-1} x^2 \tag{10}$$

$$\theta^Y = \sin^{-1} x \tag{11}$$

The target function $f(x)$ was regressed with the output of twice the Z expected value. We performed three regression tasks to verify the effectiveness of the proposed approach. A conventional least square loss function is adopted in the current regression tasks as follows:

$$L = \frac{1}{2}(2\langle Z\rangle - f(x))^2 \tag{12}$$

Moreover, its first derivation becomes:

$$\delta = \frac{\partial L}{\partial \langle Z\rangle}$$
$$= (2\langle Z\rangle - f(x)) \tag{13}$$

The error $\delta$ is the one for the backpropagation. The expectation value of $\langle Z\rangle$ is given as follows:

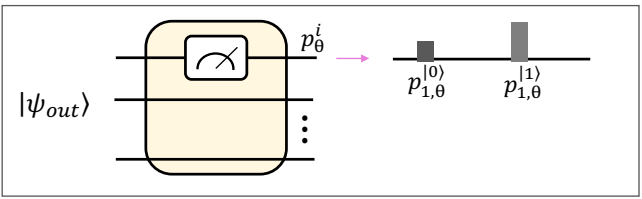

Fig. 3 Quantum circuit and measurement to obtain observation probability for regression problem.

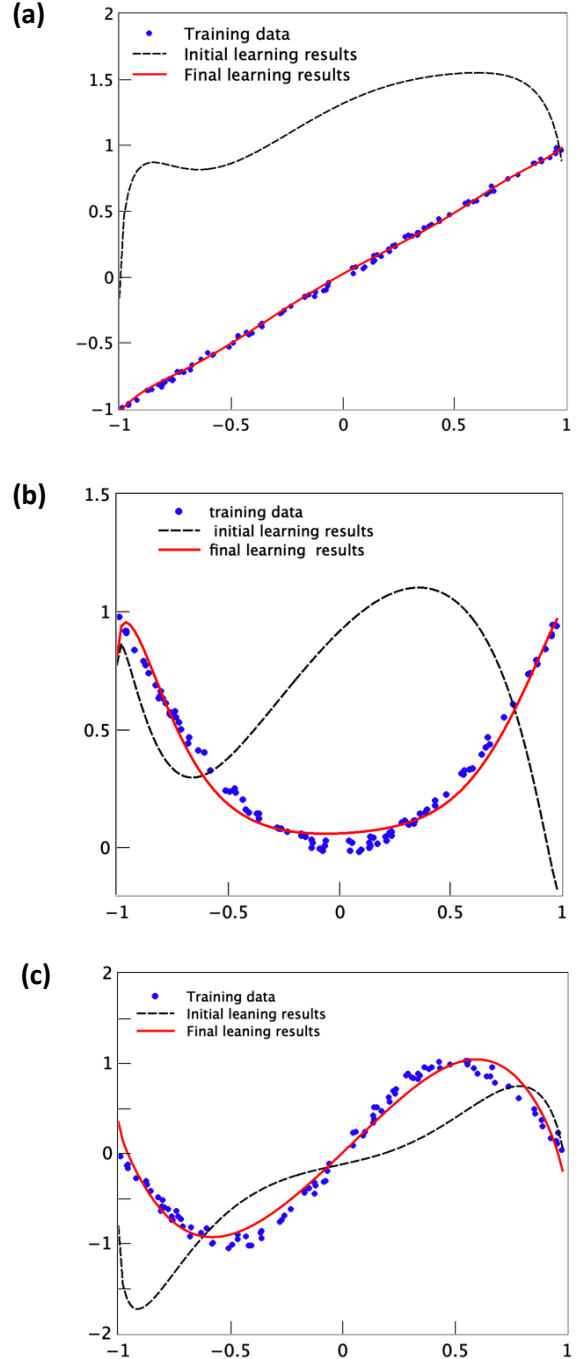

Fig. 4 (a) Regression of target function $f_1(x) = x + 0.015N(0,1)$; (b) Regression results for target function $f_2(x) = x^2 + 0.015N(0,1)$; (c) Regression results for target function $f_3(x) = \sin x + 0.015N(0,1)$,

$$\langle Z\rangle = 1 \cdot p_{1,\theta}^{|0\rangle} + (-1) \cdot p_{1,\theta}^{|1\rangle} \tag{14}$$

Here we provide with more detailed explanation regarding how the expectation value is shown in formula (14). There are two ways to obtain the probability shown in equation (14).

$p_{1,\theta}^{|i\rangle}$ can be measured through observation. For example, when we have 3 qubits quantum circuit, thus there will have a probability for eight states defined as:

$$p_\theta^{|000\rangle}, p_\theta^{|001\rangle}, p_\theta^{|010\rangle}, p_\theta^{|011\rangle}, p_\theta^{|100\rangle}, p_\theta^{|101\rangle}, p_\theta^{|110\rangle}, p_\theta^{|111\rangle}$$

If the observation measurement is performed at the first qubit, as shown in Fig.3, the probability of $p_{1,\theta}^{|0\rangle}$ and $p_{1,\theta}^{|1\rangle}$, which represent the possibility of the 1st qubit being observed as either the state of $|0\rangle$ or $|1\rangle$. The second approach to obtain the probability is by calculation using the quantum simulator. By measuring at the first qubit, $p_{1,\theta}^{|0\rangle}$ and $p_{1,\theta}^{|1\rangle}$, it implies that:

$$p_{1,\theta}^{|0\rangle} = p_\theta^{|000\rangle} + p_\theta^{|010\rangle} + p_\theta^{|100\rangle} + p_\theta^{|110\rangle} \quad (15)$$

$$p_{1,\theta}^{|1\rangle} = p_\theta^{|001\rangle} + p_\theta^{|011\rangle} + p_\theta^{|101\rangle} + p_\theta^{|111\rangle} \quad (16)$$

By doing the calculation above, the probability needed in the equation can be worked out ant thus $\langle Z \rangle$ is obtained.

Fig. 4 shows the regression results for three typical tasks to verify the validity of the proposed algorithm. In Fig.4(a), (b) and (c), three target function represent both linear and nonlinear regression was chosen as : $f_1(x) = x$, which represents a typical linear function; $f_2(x) = x^2$, which represents a single concave profile nonlinear problem, and $f_3(x) = \sin x$, which represents a multi-concave-convex wavy profile for more complex problems. The noise was also added into the target function for practical purposes, and the number of training data was chosen as 100 in circuit learning for the three target functions. It can be seen the quantum circuit based on error backpropagation performs very well in the regression task. For instance, the value of $R^2$ for the regression of $x^2$ and $\sin x$ are found as high as 0.989 and 0.992 respectively. At the initial learning stage, the results show a large deviation from the target function and at the final leaning stage, the regressed curve catches the main feature of the training data and shows the very reasonable fitted curve. It is noticed in Fig.4(a), the fitted curve showed deviation at the left edge of the regression profile. This deviation is considered as a lack of training data at the boundary and can be either improved by increasing the number of training data or adding regularization term in the loss function, which is regularly used in the conventional machine learning tasks.

*2.2 Classification*

In the classification problem, we have modified the quantum circuit architecture to accommodate the increased number of parameters for both qubit and circuit depth. The initial parameter set for classification problem was set for qubit n=4 and *l*=6 (number of layers is 7). Here we show only the results for nonlinear classification problems. The example of binary classification of the two-dimensional data is used in the experiment. Here the dataset was prepared by referring to a similar dataset from scikit-learn[27]. We consider two representative nonlinear examples: one is a dataset of make_circles, and another one is make_moons, and we consider the make_moons possess more complicated nonlinear features than make_circles. It should be noted that the data presented here are results from the sample without the addition of the noise. Due to the shortage of space, classification results for noise training data are given in the Appendix part. The two-dimensional input data $x$ was prepared by setting circuit parameters as:

$$\theta_{2i}^Z = \cos^{-1} x_1^2,$$

$$\theta_{2i}^Y = \sin^{-1} x_1 \text{ or}$$

$$\theta_{2i+1}^Z = \cos^{-1} x_2^2,$$

$$\theta_{2i+1}^Y = \sin^{-1} x_2,$$

$$(i = 0, 1) \quad (17)$$

For the training purpose, a typical cross-entropy loss function was adopted to generate the error and was further backpropagated to update the learning parameter.

$$L = d_i \log[y_1] + (1 - d_i) \log[1 - y_1] \quad (18)$$

The cross-entropy formula looks complicated but its first derivative upon the probability $y_1$ reduced to the form of error backpropagation similar to the regression tasks.

$$\delta = \frac{\partial L}{\partial \langle Z_1 \rangle} = y_1 - d_i \quad (19)$$

$$\delta = \frac{\partial L}{\partial \langle Z_2 \rangle} = -(y_1 - d_i) \quad (20)$$

For the output state $|\psi_{out}\rangle$, we calculated the expected value $\langle Z_1 \rangle$ and $\langle Z_2 \rangle$ of observable Z using the first and second qubits, Similar to the process adopted in the regression task, the final probability for the first and second qubit can be defined as follows by assuming a 3-qubit quantum circuit

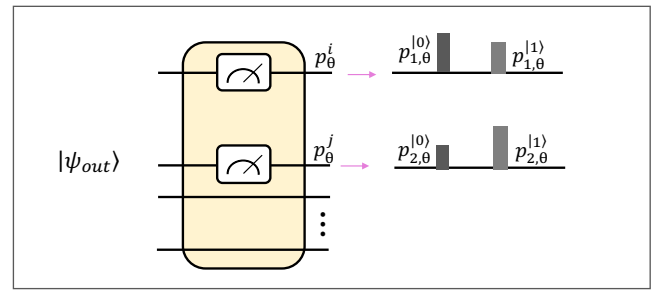

Fig. 5 Quantum circuit and measurement to obtain observation probability for classification task.

$$p_{1,\theta}^{|0\rangle} = p_\theta^{|000\rangle} + p_\theta^{|010\rangle} + p_\theta^{|100\rangle} + p_\theta^{|110\rangle} \qquad (21)$$

$$p_{1,\theta}^{|1\rangle} = p_\theta^{|001\rangle} + p_\theta^{|011\rangle} + p_\theta^{|101\rangle} + p_\theta^{|111\rangle} \qquad (22)$$

Therefore the expected value of $\langle Z_1 \rangle$ and $\langle Z_2 \rangle$ by observation measurement are given as follows:

$$\langle Z_1 \rangle = 1 \cdot p_{1,\theta}^{|0\rangle} + (-1) \cdot p_{1,\theta}^{|1\rangle} \qquad (23)$$

$$\langle Z_2 \rangle = 1 \cdot p_{2,\theta}^{|0\rangle} + (-1) \cdot p_{2,\theta}^{|1\rangle} \qquad (24)$$

Meanwhile, for the classification problem, a softmax function was applied to the output for $\langle Z_1 \rangle$ and $\langle Z_2 \rangle$ to obtain continuous probabilities $y_1$ and $y_2$ between 0 and 1. Again this treatment is the same as the method used in neural network-based machine learning classification. The obtained $y_1$ or $y_2$ can be used to calculate the loss function defined in formula (18). Here for binary classification, there exists a linear relation between $y_1$ and $y_2$ as shown in formula (25)~(27).

$$y_1 = \frac{e^{\langle Z_1 \rangle}}{e^{\langle Z_1 \rangle} + e^{\langle Z_2 \rangle}} \qquad (25)$$

$$y_2 = \frac{e^{\langle Z_1 \rangle}}{e^{\langle Z_1 \rangle} + e^{\langle Z_2 \rangle}} \qquad (26)$$

$$y_2 = 1 - y_1 \qquad (27)$$

For the proof of concept, a limited number of training data was used and was set as 200. Half of the data was labeled as '0'; the rest half of the data was labeled as '1'. For comparison, we have also applied the classical support vector machine (SVM), a toolkit attached in the scikit-learn package to the same datasets. The results from SVM are served as a rigorous reference for the validity verification of the proposed approach.

Fig.6 shows the test results for the two non-linear classification tasks. In Fig.6 (a) and Fig.6(e), 2-dimensional training data with value ranged between (-1,1) were chosen as the training dataset. Here the noise was not added for simplicity, and the training data with added noise are presented in the Appendix. Fig.6(b) shows the test results based on the learned parameter from the training dataset shown in Fig.6(a). A multicolored contour-line like classification plane was found in Fig.6(b). The multicolored value corresponding to the continuous output of the probability from the softmax function. A typical two-valued region can be easily determined by taking the median of the continuous probability as the classification boundary, and it is shown in Fig.6(b) with the line colored by pink. Reference SVM results simulated using scikit-learn-SVM is shown in Fig.6(c). Since SVM simulation treats the binary target discretely, the output shows the exact two value-based colormaps of the test results. It can be easily seen here that the results are shown in Fig.6(b) is in high consistency with the SVM results. Especially the location of the median boundary (pink line) corresponds precisely to the SVM results. For the dataset of make_moons, the situation becomes more complicated due to the increased nonlinearity in the training data. Fig.6(d), (e), and (f) showed the same simulation sequence as the data of make_circles. However, it is found both the results from error backpropagation, the approach proposed here, and SVM showed misclassification. The classification mistake usually occurs near the terminal edge area where the label '0' and label '1' overlapped with each other. Taking a closer look at the test results shown in Fig.6(e) and (f), it can be found that the misclassification presented differently. For quantum circuit learning, the misclassification occurs mostly at the left side of the label '0' in the overlapping area. For SVM, the misclassification is roughly equally distributed for both label '0' and label '1', indicating the intrinsic difference between these two simulation algorithms.

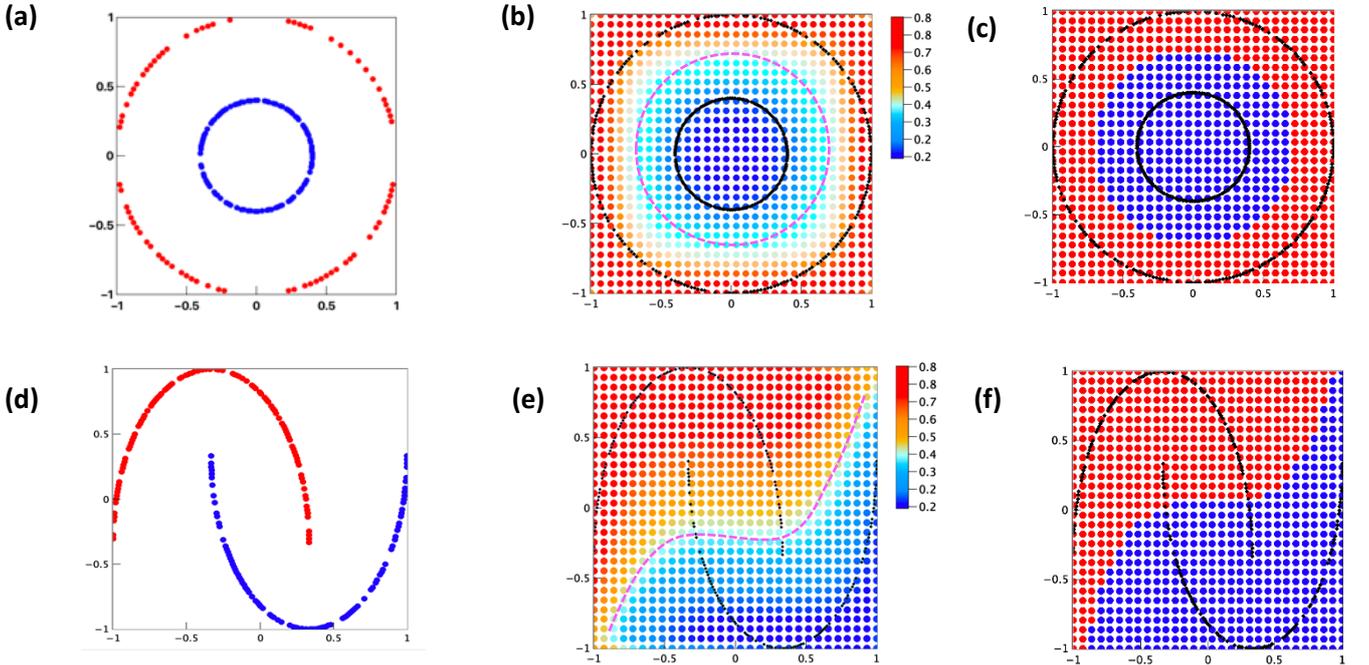

Fig. 6 Quantum circuit learning results using error backpropagation for nonlinear binary classification problem with 4 qubit and 7 layer depth: (a) Training data set for make_circles, red for label '0' and blue for label '1' ; (b) Test results using the learnt parameter using the 200 make_circles dataset, pink line corresponding to the median boundary of the continuous probability; (c) scikit-learn-SVM classification results using the learnt support vectors; (d) Training data set for make_moons, red for label '0' and blue for label '1' ;; (e) Test results using the learnt parameter under the 200 make_moon dataset, pink line corresponding to the median boundary of the continuous probability; (f) scikit-learn-SVM classification results using the learnt support vectors.

*2.3 Learning efficiency improvement*

As shown in Fig.6(d)~(f), both the backpropagation-based quantum learning algorithm and classical SVM algorithm failed to provide excellent test accuracy in the make_moon classification dataset. Further investigation aiming at improving the test accuracy for the make_moons data was conducted. Here we adopted two approaches for this purpose: (i) adjusting the depth of the quantum circuit and (ii) adjusting the scaling parameter $\gamma$ and the results are summarized as follows:

(i) Varying the depth of the quantum circuit: We consider that one of the reasons for misclassification occurred in Fig.6(e) would be attributed to the limited representation ability due to the limited depth of the quantum circuit. Therefore, we investigated the effect of quantum circuit depth on the learning accuracy and the results are shown in Fig.7(a)~(c). The depth of quantum circuit varied from 4, 7 and 10 are given. It can be seen that 4 layers of the circuit showed an almost linear separation plane, indicating the insufficient representation of the non-linear feature in the training data. However, with the increase of the circuit layer thickness, the classification boundary (separation plane) becomes more nonlinear as shown in Fig.7(b) where the depth of quantum circuit was set as 6 layers. Fig.7(c) shows the results obtained at the 10 layer depth of the quantum circuit and it can be clearly found that the separation classification plane is almost identical to the 6 layer depth ones shown in Fig.7(b), indicating the existence of a critical depth and for the depth beyond the critical depth, the learning efficiency is saturated and no further improvement could be obtained. For the current of experimental condition of 4 qubit system with 200 training dataset, the critical depth is estimated to be around 6 layers.

(ii) Varying the scaling parameter $\gamma$ : Before we present the results by varying the parameter $\gamma$, we will at first provide a detailed description about the tuning principle of $\gamma$ since it is extremely important when dealing with learning process under large scale quantum computing environment.

Parameter $\gamma$ appears in the softmax function which is used to convert the expectation values of $\langle Z_1 \rangle$ and $\langle Z_2 \rangle$ to continuous probabilities $y_1$ and $y_2$ between 0 and 1. The softmax function takes exactly the same form as shown in equation (25) and (26) except the modification shown below:

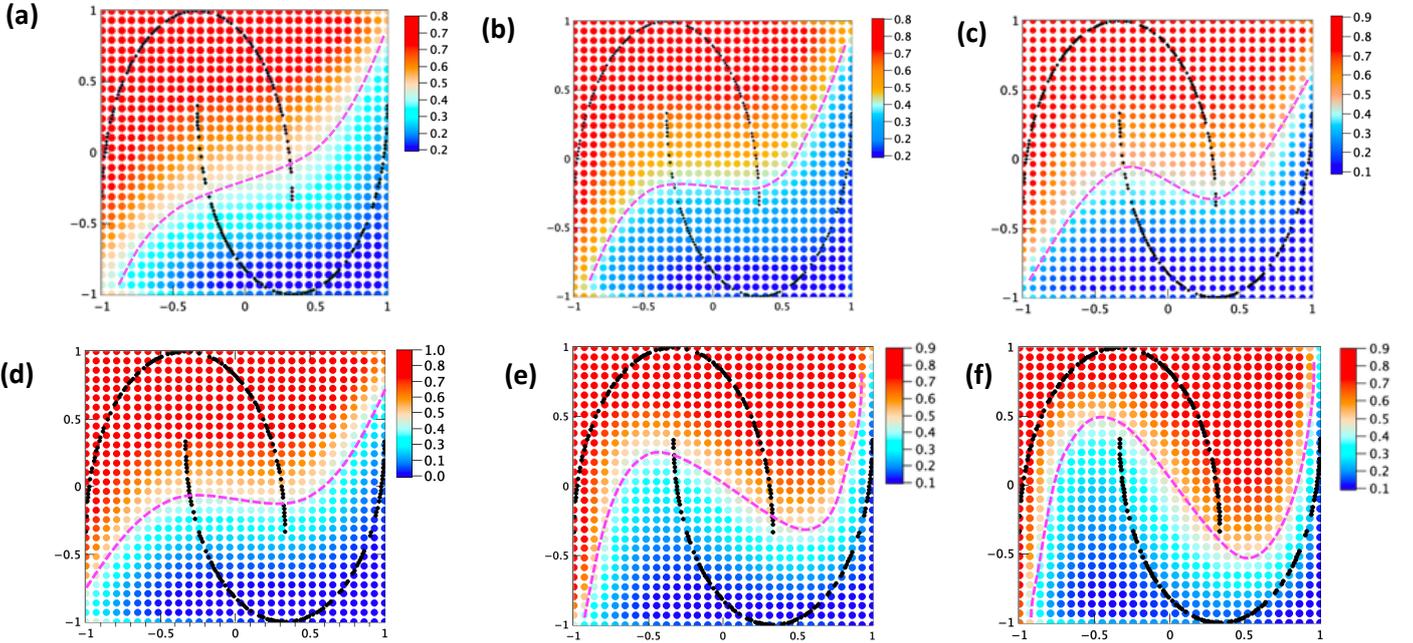

Fig. 7 Improvement of quantum learning efficiency using the 200 make_moon dataset; (i) Effect of quantum circuit depth on the classification accuracy; Training data set of label '0' and blue for label '1' are shown in dotted black line and pink line corresponds to the median boundary of the continuous probability; (a) 4 layer of the quantum circuit with 4 qubits. (b) 7 layer of the quantum circuit with 4 qubits. (c) 10 layer of the quantum circuit with 4 qubits. (ii) Effect of scaling parameter $\gamma$ on the classification accuracy: (d) $\gamma = 1$ ; (e) $\gamma = 3$ ; (f) $\gamma = 5$ .

$$y_1 = \frac{e^{\gamma\langle Z_1\rangle}}{e^{\gamma\langle Z_1\rangle} + e^{\gamma\langle Z_2\rangle}} \tag{28}$$

$$y_2 = \frac{e^{\gamma\langle Z_2\rangle}}{e^{\gamma\langle Z_1\rangle} + e^{\gamma\langle Z_2\rangle}} \tag{29}$$

In other words, as a matter of fact, in all the learning results shown so far we have assumed the parameter $\gamma = 1$. The effect of $\gamma$ on the probability value of $y$ is illustrated as follows, where we have increased the value of $\gamma$ from 1 to 3 and 5. Let as assume we have obtained two values for $\langle Z_1\rangle$ and $\langle Z_2\rangle$ as 0.3 and 0.1 respectively. The difference between these two values is very small. However, we will show that the difference between the $\langle Z_1\rangle$ and $\langle Z_2\rangle$ can be mathematically magnified by increasing the value of the parameter of $\gamma$:

$$\{\langle Z_1\rangle,\ \langle Z_2\rangle\} = \{0.3,\ 0.1\} \tag{30}$$

(1) $\gamma = 1$

$$\left\{\frac{e^{\langle Z_1\rangle}}{e^{\langle Z_1\rangle} + e^{\langle Z_2\rangle}}, \frac{e^{\langle Z_2\rangle}}{e^{\langle Z_1\rangle} + e^{\langle Z_2\rangle}}\right\} = \{0.55,\ 0.45\} \tag{31}$$

(2) $\gamma = 3$

$$\left\{\frac{e^{3\langle Z_1\rangle}}{e^{3\langle Z_1\rangle} + e^{3\langle Z_2\rangle}}, \frac{e^{3\langle Z_2\rangle}}{e^{3\langle Z_1\rangle} + e^{3\langle Z_2\rangle}}\right\} = \{0.66, 0.34\} \tag{32}$$

(3) $\gamma = 5$

$$\left\{\frac{e^{5\langle Z_1\rangle}}{e^{5\langle Z_1\rangle} + e^{5\langle Z_2\rangle}}, \frac{e^{5\langle Z_2\rangle}}{e^{5\langle Z_1\rangle} + e^{5\langle Z_2\rangle}}\right\} = \{0.73, 0.27\} \tag{33}$$

As shown in formula (31)~(33), an increase of the parameter $\gamma$ has significantly enhanced the difference between the converted probability $y$. The enlarged difference is expected to improve the learning efficiency in the classification problem by quickly determining the separation plane between the binary training data.

To verify the effect from the scaling parameter $\gamma$, we performed a further experiment on the make_moon data. The results by tuning scaling parameter $\gamma$ are summarized in Fig.7(d)~(f) showing the results from three cases regarding $\gamma = 1,\ \gamma = 3, \gamma = 5$. In all the experiments, the number of qubits was kept at 4 qubits. It can be found that the scaling parameter $\gamma$ exerts a significant effect on learning efficiency. The classification accuracy is dramatically improved when $\gamma$ is set to 5 shown in Fig.7(f). By checking the contour separation line shown in Fig.7(f), it can be quickly confirmed that the classification accuracy has reached almost 100%, indicating the effectiveness of scaling parameter $\gamma$ in improving leaning efficiency. It is also worthwhile to mention here that the probability of each quantum states has to be normalized to ensure the summation $\sum p_i = 1$. This constraint strongly suppresses the probability of each state and the final probability difference between each state at the initial leaning stage tends to become extremely small due to the exponential increase of the representation state as $2^{N_{qubit}}$ at the large-scale quantum computing system. We claim that it is crucial to tune the scaling parameter $\gamma$ at large scale qubit involved NISQ system for good leaning performance.

## 2.4 Computation efficiency

After having confirmed the validity of the proposed error backpropagation on various regression and classification problems, we will show one great advantage of using backpropagation to perform parameter optimization over other approaches. It has been rigorously demonstrated in a deep neural network-based machine learning field that the error backpropagation method shows several magnitudes faster than the conventional finite difference method in gradient descent-based learning algorithms. In this work, we conducted a benchmark test to verify where there is a decisive advantage of using a backpropagation algorithm in quantum circuit learning. Fig.8 shows the computation cost comparison among three methods: a finite difference method proposed in reference [22], the popular SPSA method which is currently used in complicated quantum circuit leaning [26], and the proposed method based on backpropagation. The execution time with the unit of a second per 100 iterations is selected for a fair comparison. The number of parameters corresponding to the quantum circuit depth $l$ and number of qubits $O_{qubit}$ is given as:

$$N_{parameters} = (S_{rotation-gate}) \times (O_{qubit}) \times (l+1) \quad (34)$$

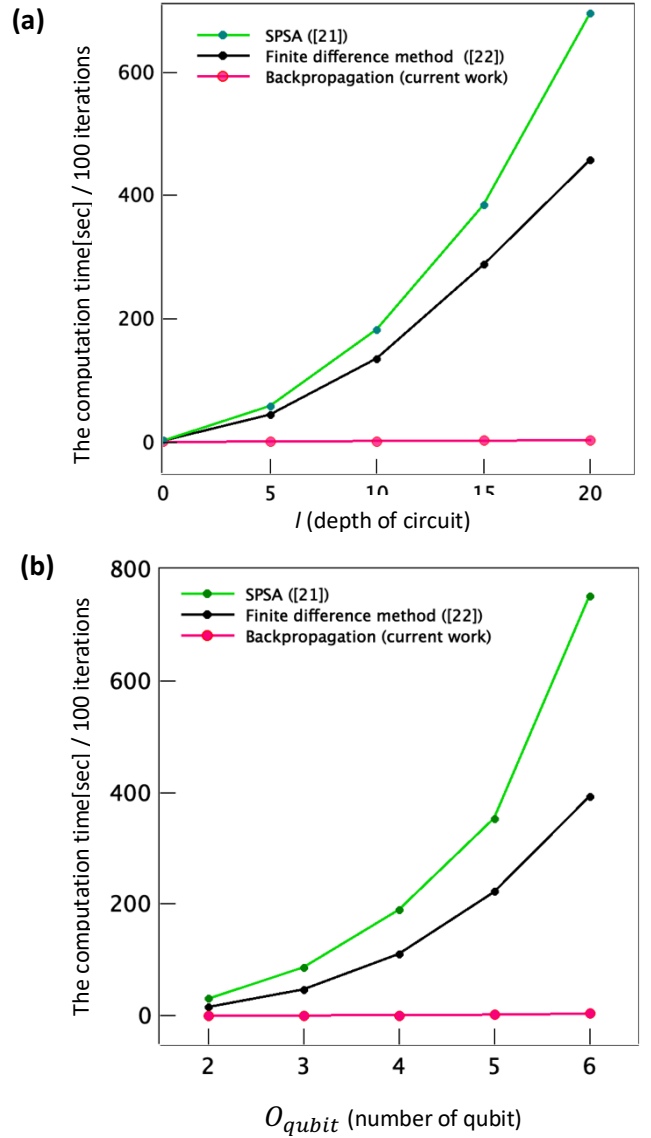

Fig. 8 Comparison of computation cost for different approaches. (a) Computation cost dependence on the depth of the quantum circuit; (b) Computation cost dependence on the number of qubits.

The result of the comparison by both varying the depth of the quantum circuit and the number of qubits is presented in Fig.8. We implemented the three methods on the same make_moons dataset and recorded the computation time cost per 100 iterations. Fig.8(a) shows the dependence of computation cost on the variation of depth of the quantum circuit. In this experiment, we have fixed the number of quantum bit $O_{qubit}$ as 4 qubits. The depth of the quantum circuit is varied from 5 to 20 at the interval of 5. It can be seen there is a dramatic difference in computation time cost for 100 iteration learning steps. The finite difference method and the SPSA method showed poor computation efficiency, as has been mentioned above and demonstrated in the deep neural network-related machine learning field. The computation costs rise exponentially as the thickness of the circuit increases, limiting its application in the large scale and deep quantum circuit. In contrast, the backpropagation method proposed here showed a dramatic advantage over all other methods by shown an almost constant dependence on the depth of the quantum circuit. The computation time is recorded at a depth of 20 layers as 3.2 seconds, which is almost negligible when compared to the recorded value at the same 20 layer thickness: 458 seconds by using the finite difference method and 696 seconds by using SPSA method.

Fig.8(b) shows the dependence of computation cost on the variation of the number of qubits. In this experiment, we have fixed the number of depths of the quantum circuit as ten layers. The number of qubits is varied from 2 to 6 at the interval of 1. Similar to the tendency found in Fig. 8(a), there is a dramatic difference in computation time cost for 100 iteration learning steps. The finite difference method and the SPSA method showed poor computation efficiency, and the profile is similar to those shown in Fig.8(a). The computation costs rise exponentially as the $O_{qubit}$ Increases, limiting its application in the large scale and deep quantum circuit. In contrast, the backpropagation method proposed here showed a dramatic advantage over all other methods by shown an almost constant dependency on the $O_{qubit}$. The computation time recorded at the number of qubits as 6 is around 4.1 seconds, which is almost negligible when compared to the recorded value at the same six qubits as 393 seconds by using the finite difference method and 752 seconds by using SPSA method.

## 3  Discussion

So far, we have focused our results on the simulation using the quantum simulator. Implementing architecture when using a real machine such as NISQ type quantum is described in Fig.9. To use the error backpropagation method, a quantum state $|\psi\rangle$ is required to get prepared at the initial stage. Therefore, as shown in the figure, a quantum circuit having the same configuration as the real quantum circuit must be prepared as a quantum simulator on a classical computer. It should be noticed here that this could not be considered as additional load for the quantum computing scientist since a quantum computer is not allowed to be disturbed during the working condition, unlike the classical computer, it needs its counterpart of quantum circuit simulator to monitor and diagnose the qubits and gate error and characterizing the advantage of quantum computers over classical computers [28–33], Therefore, for a real quantum computer, it always requires a quantum simulator ready for use at any time. That means we can always access the quantum simulator, as shown on the right side of Fig.9, to examine and obtain detailed information regarding the performance of the corresponding real quantum computer. Observation probability for each state $|\psi_j\rangle$ can be calculated by shooting $R$ times at the real quantum computer side. The observation probability obtained from the real quantum machine is then passed to the classical computer, and the quantum circuit in the simulator for simulation is then used. The parameter $\theta$ can be updated using backpropagation since all the intermediate information is available at the simulator side. After the parameter $\theta^*$ is updated at the simulate side, it will return to the real quantum machine for next iteration quantum simulation. Implementing the backpropagation will be reported elsewhere. There may raise a concern about the feasibility of the proposed approach on the

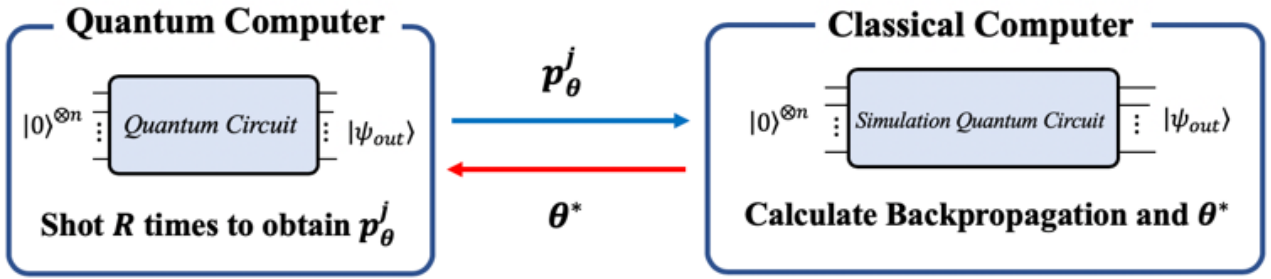

Fig.  9 Implementation architecture of error backpropagation-based quantum circuit learning on the real NISQ type quantum computer

quantum circuit with hundreds or thousands number of qubits

$N_{qubit}$. We indeed need a storage capacity of $2^{N_{qubit}}$ to accommodate all the states in order to perform the error backpropagation well and it turns out to be extremely challenging when $N_{qubit}$ is very large. For an 'authentic' quantum algorithm, the algorithm may indeed have been designed in a way that we do not need $2^{N_{qubit}}$ memory to record all the states because most of the amplitude of the state has vanished during the quantum operation. The word 'authentic' implies a complete end-to-end quantum algorithm. However, as mentioned in the reference [28-33], quantum computing and algorithm design must be guided by an understanding of what tasks we can hope to perform. This means that an efficient, scalable quantum simulator is always vital for the 'authentic' quantum algorithm. Since the error-backpropagation is performed layer by layer over matrix operation, more advanced GPGPU based algorithm, tensor contraction or the path integral based sum-over-histories method would be effectively used to tackle the $2^{N_{qubit}}$ operation[34–39]. Therefore, the concern raised above will be relieved or eliminated over the improvement of the quantum computing field and GPGPU field as well as other surrounding techniques.

## 4 Conclusion

We proposed a backpropagation algorithm for quantum circuit learning. The proposed algorithm showed success in both linear and nonlinear regression and classification problems. Meanwhile, dramatic computation efficiency by using the error backpropagation based gradient circuit learning rather than the other gradient-based method such as Finite difference method or SPSA method. The reduction of computing time is surprisingly up to several magnitudes when compared to the conventional method. Further computing advantage would be expected by combining the backpropagation with the GPGPU technique. Therefore, we claim that gradient descent using the error-backpropagation could be used as an efficient quantum circuit learning tool both in the NISQ era and still be useful even in a more matured quantum computer with a large scale of circuits depth and thousands of quantum bit.

ACKNOWLEDGMENT

This work is supported by New Energy and Industrial Technology Development Organization (NEDO), and Ministry of Economy, Trade and Industry (METI), Japan. We also appreciate Dr.Chih-Chieh Chen for his inspiring suggestions and comments.